\documentclass[%
 reprint,
 amsmath,amssymb,
 aps
,prc,
]{revtex4-2}
\usepackage{graphicx}
\usepackage{dcolumn}
\usepackage{bm}
\usepackage{natbib}
\usepackage{color}

\usepackage{hyperref}
\usepackage{cleveref}
\newcommand{\bs}[1]{\boldsymbol{#1}}

\newcommand{\rint}[1]{\int \frac{\mathrm{d}^4{#1}}{(2\pi)^4}}
\begin{document}

\preprint{APS/123-QED}

\title{Tunnelling amplitudes through localised external potentials from Feynman diagram summation}

\author{Rosemary Zielinski$^{1}$}
\email{rosemary.zielinski@anu.edu.au}
\author{C\'edric Simenel$^{1,2}$}%
\email{cedric.simenel@anu.edu.au}
\author{Patrick McGlynn$^{1,2}$}

\affiliation{%
 $^{1}$Department of Fundamental and Theoretical Physics, The Australian National University
}%

\affiliation{%
 $^{2}$Department of Nuclear Physics and Accelerator Applications, The Australian National University
}%

\date{\today}

\begin{abstract}
{Currently there is no general theory of quantum tunnelling of a particle through a potential barrier which is compatible with QFT. We present a complete calculation of tunnelling amplitudes for a scalar field for some simple potentials using quantum field-theoretic methods. Using the perturbative $S$-matrix formalism, starting with the Klein-Gordon Lagrangian, we show that an infinite summation of Feynman diagrams can recover tunnelling amplitudes consistent with relativistic quantum mechanics. While this work does not include many-particle effects arising from a fully quantised QFT, it is necessary to investigate QFT corrections to tunnelling amplitudes. 
}
\end{abstract}

\maketitle
{Quantum tunnelling is the phenomenon where particles probabilistically traverse regions energetically forbidden by classical mechanics. Despite its widespread technological applications including tunnel diodes \cite{Esaki_1958}, 
scanning tunnelling microscopy \cite{binnig_scanning_1987, binnig_scanning_1983}, 
as well as its importance to many biological \cite{Devault_1980,trixler_quantum_2013} 
and chemical systems, the theoretical foundations of quantum tunnelling are poorly understood.}
While single-particle quantum tunnelling can be studied in the context of relativistic quantum mechanics with the Klein-Gordon equation (spin~0) or with the Dirac equation (spin~$\frac{1}{2}$), or with  the non-relativistic Schr\"odinger equation, there is no comprehensive description of quantum tunnelling using quantum field theory (QFT). Though the notion of quantum tunnelling between configurations of a field is established with {\it instanton} methods \cite{coleman_uses_1979} and in the context of false vacuum decay theories \cite{coleman_fate_1977,coleman_1977_2, devoto_false_2022}, this work addresses the more conventional idea of quantum tunnelling of a particle through a localised \textit{external potential}. 

Unlike quantum mechanics, where some interactions are described by non-local phenomenological potentials (such as the Coulomb interaction), interactions in QFT are described by local couplings between quantised fields. This property gives rise to many-particle effects beyond the scope of single-particle quantum mechanics, including pair-production and virtual-particle mediators. These many-particle effects may have important consequences for quantum tunnelling. The archetypal example is the Klein paradox \cite{Klein_1929}, where fermions incident on a step potential of height $V$ exhibit an apparent violation of unitarity when $V$ is greater than the Schwinger limit, $eV>2m$. This paradox is not a consequence of a non-relativistic treatment, given that the initial calculation is fully in accordance with relativistic quantum mechanics. The canonical resolution, presented in \cite{hansen_kleins_1981}, accounts for the lack of unitarity via a consideration of pair-production at the barrier, thus invoking QFT. More recent treatments of  critical potential steps \cite{gavrilov_2020,gavrilov_2016} carefully consider the in- and out- particle operators for time-independent potentials using solutions to the Dirac equation. The key challenge is relating the initial particle vacuum to the final state vacuum, which are no longer equivalent with a step potential due to the non-vanishing external background inducing pair production. This relationship can be determined for analytic potentials such as the Klein step or Sauter potential \cite{kim_2010,hansen_kleins_1981}, though it is not known for general potentials. Besides the Klein paradox, there are many reasons to consider QFT: it is our most fundamental theory of physics, with unrivalled success in predicting the standard model, explaining the anomalous magnetic moment \cite{schwinger_quantum-electrodynamics_1948,aoyama_revised_2018}, and accounting for hyperfine electronic structure such as the Lamb shift \cite{lamb_fine_1947,bethe_electromagnetic_1947}.  

Though there is a clear motivation to integrate the theory of quantum tunnelling with QFT, the computational methods within QFT present challenges in this pursuit. In particular, QFT is often formulated with path-integral  or canonical quantisation methods, both of which cannot be solved exactly for any meaningful interacting theory in four dimensions. In practice, most physical QFT calculations are obtained via perturbative methods. One way to relate initial states to final states in QFT uses the scattering matrix ($S$-matrix), a unitary time evolution operator. In the interaction picture, it has the form 
\begin{align}
S &= \mathcal{T}\left [\mathrm{e}^{-i\int \mathrm{d}^4x H_{int}(x)}\right ],
\end{align}
where $\mathcal{T}$ is the time-ordering operator, and $H_{int}(x)$ is the interaction Hamiltonian. The time-ordering operator makes this cumbersome to compute exactly, and so a truncated Dyson series expansion is often employed,
\begin{widetext}
\small
\begin{align}
S = \sum _{n=0}^{\infty} \frac{(-i)^n}{n!}\int \mathrm{d}^4x_1\int \mathrm{d}^4x_2 \dots \int \mathrm{d}^4x_n \mathcal{T}\left [H_{int}(x_1)H_{int}(x_2)\dots H_{int}(x_n)\right ].
\end{align}
\end{widetext}
\normalsize
Each term in the expansion corresponds to a set of Feynman diagrams at the given order in the interaction Hamiltonian. This perturbative approach is remarkably powerful, especially for quantum electrodynamics (QED) where the fine structure constant is small. However, perturbative methods break down in the presence of strong electromagnetic fields, famously exemplified by the non-perturbative nature of the Schwinger effect \cite{schwinger_gauge_1951}. For a discussion of other strong-field non-perturbative effects, see \cite{fedotov_advances_2023}. 

Quantum tunnelling is another fundamentally non-perturbative phenomenon: by definition, any interactions with the external barrier cannot be considered small in the tunnelling regime. It is this fact which makes a QFT treatment of quantum tunnelling difficult, and resistant to typical finite-order $S$-matrix calculations. Another issue arises from the nature of the $S$-matrix itself: by construction, it describes the evolution of asymptotically free initial states to asymptotically free final states. 
This precludes any bound-state tunnelling calculations, such as those associated with nuclear fusion. However, the $S$-matrix formalism still has the potential to describe tunnelling through barriers of finite width, where the initial and final states are described by free particles.

While tunnelling of a particle through external potentials has, to our knowledge, never been explicitly considered within a QFT framework, there exist formalisms to calculate above-barrier scattering from external potentials in both QFT and relativistic quantum mechanics (RQM). Electron scattering from external potentials is treated both with the path integral formalism in \cite{Xu_2016} and with the canonical quantisation formalism in \cite{de_leo_2009}, where both papers recover identical perturbative scattering amplitudes. Though each work uses QFT \textit{methods}, they both begin with the Dirac Lagrangian with a classical external field, with the absence of a quantised electromagnetic field. In this case, the Lagrangian describes a single-particle RQM theory, rather than a full QFT. The present work seeks to extend the single-particle scattering calculations presented in \cite{Xu_2016, de_leo_2009} into the tunnelling regime, to demonstrate a quantum field-theoretic approach to determining tunnelling amplitudes. Rather than consider particle spin, this work focuses on scalar fields as the simplest proof-of-concept of field-theoretic quantum tunnelling. We start with the Klein-Gordon equation with a scalar external potential, equivalent to a mass perturbation, and demonstrate that tunnelling amplitudes congruent with RQM can be found using an infinite summation of Feynman diagrams. The potentials are chosen to be one or two Dirac delta functions. Though this work does not consider the many-particle effects intrinsic to a full QFT treatment of tunnelling, a demonstration that tunnelling amplitudes can be obtained with QFT methods is the necessary precursor to investigating the effect of quantum corrections. 
 \subsection*{Formalism \label{sec:formalism}}
This work considers the tunnelling/scattering behaviour of a neutral scalar field $\phi(x)$ in the presence of an external potential, $u(x)$. The relevant Lagrangian is given by 
\begin{align}
  \mathcal{L} &= \frac{1}{2}\partial_{\mu}\phi\partial^{\mu}\phi-\frac{1}{2}m^2\phi^2-\frac{1}{2}e u(x)\phi^2,
  \label{eqn:KG_lagrangian}
\end{align}
where $m$ is the particle mass and $e$ is a coupling constant. We stress that $u(x)$ is not a dynamical field. This leads to the Klein-Gordon equation with a linear source, 
\begin{align}
  (\partial^2 +m^2)\phi &= -e u(x)\phi. \label{eqn:KG_eqn}
\end{align}

Rather than solve this system using methods in partial differential equations (PDEs), this section demonstrates how tunnelling amplitudes may be obtained via a series of Feynman diagrams. We note $p$ is the incoming four-momentum and $k$ the outgoing one. The Feynman rules in momentum space associated with the Lagrangian~(\ref{eqn:KG_lagrangian}) are:
\begin{enumerate}
  \item Each vertex produces a 
  factor  $-ie \tilde{u} (p-k)$, with $\tilde{u}(q) = \int \mathrm{d}^4 x \,u(x)\mathrm{e}^{iqx}$. 
  \item Internal lines have a propagator, $D(q) = \frac{i}{q^2-m^2+i\varepsilon}$.
  \item Momenta of internal lines are integrated over. 
\end{enumerate}
The $S$-matrix expansion for a neutral scalar field interacting with the external field $u(x)$ has the diagrammatic expansion
\begin{align}
  \langle k | S | p \rangle &= \vcenter{\hbox{\includegraphics[scale=0.4]{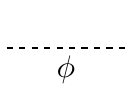}}} + \vcenter{\hbox{\includegraphics[scale=0.4]{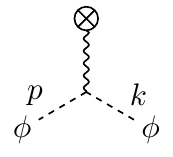}}}+\vcenter{\hbox{\includegraphics[scale=0.4]{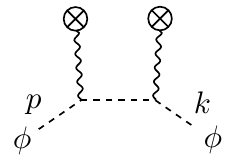}}}+\vcenter{\hbox{\includegraphics[scale=0.4]{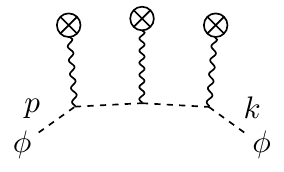}}}\nonumber\\
  &= \sum_{n=0}^{\infty} \langle k | S | p\rangle^{(n)} \label{eqn:diagram_expansion}
\end{align}
where $\bigotimes$ represents the static, external field. The number of vertices on each Feynman diagram denotes the order of interaction, and $\langle k |S | p \rangle ^{(n)}$ the $n^{\text{th}}$-order contribution to the $S$-matrix. Here, $|k\rangle$ and $|p\rangle$ represent single-particle momentum eigenstates. Though $\langle k | S | p \rangle $ denote $S$-matrix elements for single-particle initial and final states, there is no global momentum conservation as the external potential  breaks translational  invariance. Additionally, no loop diagrams can be generated with the Lagrangian in Eq.~(\ref{eqn:KG_lagrangian}), because it contains only the dynamical field $\phi$, which only interacts with the external field (i.e., there are no self-interactions).
 
 Let us start with the zeroth-order term in Eq.~(\ref{eqn:diagram_expansion}): 
\begin{align}
  \langle k | S |p \rangle ^{(0)} = (2\pi)^3 2E(\bs{p}) \delta^3(\bs{p}-\bs{k}).\label{eqn:S_0}
\end{align}
Here $E(\bs{p}) = |\bs{p}|^2 + m^2$, where $\bs{p}$ is the Euclidean momentum of the particle. 
This non-interacting term 
is the identity when integrated over Lorentz-invariant phase space. Though this term is typically neglected for cross-section calculations, we retain it to preserve unitarity in final reflection and transmission coefficients. 

The first-order  $S$-matrix element, given by
\begin{align}
  \langle k | S | p \rangle ^{(1)} = \vcenter{\hbox{\includegraphics*[scale=0.4]{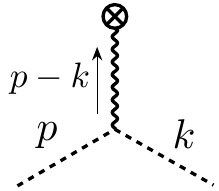}}}= -ie\tilde{u}(p-k), \label{eqn:S_1}
\end{align}
recovers the Born approximation. 
In fact, higher order Feynman diagrams are more interesting as they include internal lines. For instance, the second-order term reads
\begin{align}
  \langle k | S | p \rangle ^{(2)} &=  \vcenter{\hbox{\includegraphics*[scale=0.4]{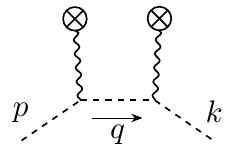}}}\nonumber\\
&= (-ie)^2 \int \frac{\mathrm{d}^4 q}{(2\pi)^4}\frac{i\tilde{u}(p-q)\tilde{u}(q-k)}{q^2-m^2+i\varepsilon}.\label{eqn:S_2}
\end{align}
Unlike typical Feynman diagrams which only contain \textit{dynamical} fields, the scalar field $\phi$ can exchange an arbitrary amount of momentum with the external field $u$, and thus the momentum $q$ of the internal line is not specified. 

Generalising to the $n^{\text{th}}$-order, 
\begin{widetext}
\begin{align}
  \langle k | S | p \rangle ^{(n)} &= (-ie)^n \left(\prod^{n-1}_{j=1}\rint{q_{j}}\,\prod_{i=1}^{n-1} D(q_i)
\prod_{k=1}^{n-2}\tilde{u}(q_k-q_{k+1})\right)\tilde{u}(p-q_1)\tilde{u}(q_{n-1}-k)\nonumber\\
  &= -ie \rint{q} \frac{i\tilde{u}(q-k)}{q^2-m^2+i\varepsilon}\langle q | S | p \rangle ^{(n-1)}, \label{eqn:Sn_recursive}
\end{align}
\end{widetext}
from which we note that $n-1$ integrals over internal lines are required. 
The final step in Eq.~(\ref{eqn:Sn_recursive}) allows a definition in terms of the previous $S$-matrix element. Thus, the total $S$-matrix element can be written as
\begin{align}
  \langle k | S | p \rangle &= \langle k | S |p \rangle^{(0)} + \rint{q}G_{\phi}(q-k)\langle q | S | p \rangle ^{(1)},
\end{align}
with $G_{\phi}(q-k)$ the Green's function satisfying the Fredholm integral equation
\begin{align}
  (2\pi)^4 \delta^4(p-k) &= G_{\phi}(p-k) \nonumber\\
  &+ie \rint{q}\frac{i\tilde{u}(q-k)}{q^2-m^2+i\varepsilon} G_{\phi}(p-q). \label{eq:Fredholm}
\end{align}
Due to the non-perturbative nature of tunnelling, an analytic solution is required. 
However, there is no general solution to Eq.~(\ref{eq:Fredholm}) for arbitrary potentials. 
In order to progress, we consider the simplest case for which the sum in Eq.~(\ref{eqn:diagram_expansion}) can be computed analytically, namely a Dirac delta potential. 

\subsection*{Tunnelling through 1D Dirac delta potentials \label{sec:delta}}
The delta potential is a limit of the traditional rectangular barrier with width $a\rightarrow0$ and height $V_0\rightarrow\infty$.
We define a one-dimensional static barrier 
\begin{align}
  u(x) = a V_0 \delta(x_3)
\end{align}
giving the four-dimensional Fourier transform 
\begin{align}
  \tilde{u}(p-k) = (2\pi)^3 aV_0 \prod_{\mu=0}^{2}\delta(p_\mu-k_\mu).
\end{align}
Using this definition in Eqs.~(\ref{eqn:S_1}) and~(\ref{eqn:S_2}) gives
\begin{align}
  \langle k | S | p \rangle ^{(1)} &= -ieaV_0 (2\pi)^3\prod_{\mu=0}^{2}\delta(p_\mu-k_\mu), \label{eq:Sdelta1}
\end{align}
and
\small
\begin{align}
  \langle k | S | p \rangle ^{(2)} &= \rint{q}\frac{i(-ieaV_0)^2 (2\pi)^6}{q^2-m^2+i\varepsilon} \prod_{\mu=0}^{2}\delta(p_\mu-q_\mu)\delta(q_\mu-k_\mu)\nonumber\\
  &= (-ieaV_0)^2(2\pi)^2\prod_{\mu=0}^{2}\delta(p_\mu-k_\mu)\nonumber\\
  &\times \int \frac{i\mathrm{d}q_3}{p_0^2-p_1^2-p_2^2-q_3^2-m^2+i\varepsilon}.
\end{align}
\normalsize

The particle is on mass-shell in the initial and final states.
Considering the case of an incident particle originating at $x_3\rightarrow-\infty$, i.e., with $p_3>0$, this condition gives $p_3=\sqrt{p_0^2-p_1^2-p_2^2-m^2}$.
Applying the residue theorem allows the computation of the final integral over $q_3$:
\begin{align}
  \langle k | S | p \rangle ^{(2)} &= \frac{(-ieaV_0)^2}{2 p_3}(2\pi)^3 \prod_{\mu=0}^{i=2}\delta(p_{\mu}-k_{\mu})\nonumber\\
  &= \frac{-ieaV_0}{2p_3}\langle k | S | p \rangle^{(1)}.\label{eq:Sdelta2}
\end{align} 
We see from Eq.~(\ref{eq:Sdelta1}) that $\langle q|S|p\rangle^{(1)}$ does not depend on $q_3$. 
The same naturally applies to $\langle q|S|p\rangle^{(2)}$ in Eq.~(\ref{eq:Sdelta2}). 
This follows from the fact that Dirac delta potential does not provide any natural length scale, and thus it has no momentum dependence after Fourier transform. 
This generalises to all orders.
Thus, from Eq.~(\ref{eqn:Sn_recursive}), we see that each $n^{\text{th}}$-order term of the $S$-matrix can be expressed as
\begin{align}
  \langle k | S | p \rangle ^{(n)} &= \frac{-iaeV_0}{2p_3} \langle k | S |p \rangle ^{(n-1)}\\
  &=  \left(\frac{-iaeV_0}{2p_3}\right)^{n-1} \langle k | S |p \rangle ^{(1)}
\end{align}
for $n \geq 1$. This $S$-matrix expansion therefore forms a geometric series. In the perturbative regime ($2p_3 > |aeV_0|$), this series is convergent, and the sum in Eq.~(\ref{eqn:diagram_expansion}) can be done analytically, giving
\begin{align}
  \langle k | S | p \rangle = \langle k | S | p \rangle ^{(0)} + \frac{1}{1+\frac{iaeV_0}{2p_3}}\langle k | S | p \rangle ^{(1)}. \label{eqn:S_total}
\end{align}


It remains to extract transmission and reflection coefficients from the $S$-matrix elements. Following the procedure in Ref.~\cite{de_leo_2009}, the transmission coefficient is simply the integral of the $z$-component of the final momentum ($k_3$) over the positive real domain:
\begin{align}
  T = \int_{-\infty}^{\infty}\mathrm{d}k_1 \int_{-\infty}^{\infty}\mathrm{d}k_2 \int_{0}^{\infty}\frac{\mathrm{d}k_3}{(2\pi)^3 2 E (\bs{k})} \langle k | S | p \rangle.\label{eqn:trans_defn}
\end{align}
Inserting Eq.~(\ref{eqn:S_total}) into Eq.~(\ref{eqn:trans_defn}), and using Eq.~(\ref{eqn:S_0}), we obtain
\begin{align}
  T &= 1 + \int_{0}^{\infty} \frac{\mathrm{d}k_3}{2E(\bs{k})}\frac{-iaeV_0}{1+\frac{iaeV_0}{2p_3}}\delta(p_0-k_0)
\end{align}
where the two integrals over $k_1$ and $k_2$ are performed trivially. To compute the remaining integral, we consider an incoming particle, with initial momentum $p = (p_0,0,0,p_3\!\!>\!\!0)$. In this case, $p_0^2 = p_3^2+m^2$, and $\delta(p_0-k_0) = \frac{p_0}{p_3}(\delta(p_3-k_3)+\delta(p_3+k_3))$. This leads to
\begin{align}
  T &= \frac{1}{1+\frac{iaeV_0}{2p_3}}.
\end{align}
A similar calculation for the reflection coefficient yields
\begin{align}
  R &= \int_{-\infty}^{\infty}\mathrm{d}k_1 \int_{-\infty}^{\infty}\mathrm{d}k_2 \int_{-\infty}^{0}\frac{\mathrm{d}k_3}{(2\pi)^3 2 E (\bs{k})} \langle k | S | p \rangle \nonumber\\
  &= \frac{1}{\frac{2ip_3}{aeV_0}-1}.\label{eqn:ref_defn}
\end{align}
We note that $|R|^2 + |T|^2 = 1$, as desired. Additionally, these transmission and reflection coefficients exactly coincide with those derived starting from the Klein-Gordon equation and imposing the usual boundary conditions (see Methods). 

We emphasise that the total $S$-matrix element in Eq.~(\ref{eqn:S_total}) has been computed in the perturbative regime, where $2p_3 \geq |eaV_0|$. Outside this radius of convergence, the series representation defined by Eq.~(\ref{eqn:Sn_recursive}) is no longer equal to Eq.~(\ref{eqn:S_total}). This process has therefore analytically continued the series beyond its original domain in the quantity $p_3$, in order to capture the non-perturbative regime and thus tunnelling behaviour.


A single-delta potential is an important limit of a rectangular barrier, and does approximate systems such as two conductors with a  small insulating gap. However, by construction, there is no associated length scale. Thus, tunnelling through such a potential cannot feature resonance and interference effects so often associated with tunnelling. One simple way to introduce a length scale is with two delta potentials, spaced with some distance $b$ 
\begin{align}
u(x) &= \frac{aV_0}{2}\left [\delta(x_3 + \frac{b}{2}) + \delta(x_3 - \frac{b}{2})\right ]\label{eq:pot2delta}
\end{align}
with the factor $\frac{aV_0}{2}$ chosen such that the integral of the spatial potential remains the same as the single-delta case. 

Repeating the above derivation for the two-delta potential (see Methods) leads to the transmission and reflexion coefficients: 
  \begin{align}
    T_{2 \delta}  =& \frac{16 p_3^2 }{(a e V_0e^{i b p_3})^2+ (4 p_3+ ia e V_0)^2}\label{eq:T2}\\
    R_{2 \delta} =& -\frac{2 i a e V_0 [a e V_0 \sin (b p_3)+4 p_3 \cos (b p_3)]}{(a e V_0e^{i b p_3})^2+ (4 p_3+ia e V_0)^2}.\label{eq:R2}
    \end{align}
As in the case of a single delta-function, these amplitudes exactly match the standard amplitudes found using relativistic quantum mechanics (see Methods) and obey $|T_{2\delta}|^2+|R_{2\delta}|^2=1$. 
 
\subsection*{Discussions and Conclusions}
The validity of our tunnelling results rests entirely on the process of analytic continuation, from above-barrier scattering to the entire momentum regime. As pointed out by \cite{de_leo_2009}, this analytic continuation breaks down in the Klein regime, where pair-production begins. However, for the case of a \textit{neutral scalar} field, the Klein paradox never arises, simply because there are no anti-particles. Therefore, this infinite Feynman diagram summation technique, at least in this system, will always yield results which can be analytically continued to the tunnelling regime. Indeed, this is also true of other systems, such as a complex scalar field $\phi\ne\phi^\dagger$ coupled to an ungauged potential term, acting as a mass perturbation, 
\begin{align}
\mathcal{L}=|\partial_{\mu}\phi|^2-m^2|\phi|^2-eu|\phi|^2,
\end{align}
which yields an identical Klein-Gordon equation to Eq.~(\ref{eqn:KG_eqn}), despite the presence of antiparticles. 
In this case, the potential acts as an effective mass affecting both particles and antiparticles in the same way.
In contrast, a complex scalar coupled to a vector potential $A_{\mu}(x)$ via
\begin{align}
\mathcal{L}= |(\partial_{\mu}-ieA_{\mu})\phi|^2-m^2|\phi|^2
\end{align}
will encounter the Klein regime depending on the strength of the potential, as will spinor field obey Dirac equation coupled to the same vector potential. The issues with the Klein regime further highlight the difficulty associated with tunnelling in QFT: while we have demonstrated that a QFT treatment of tunnelling is possible within a single-particle RQM framework, this approach fails to capture the quintessential many-particle nature of QFT tunnelling. It leaves the open question of how to properly address QFT corrections to tunnelling, potentially providing a rich starting point for future work. However, this work is a valuable proof-of-concept which relates standard calculation techniques in QFT, such as Feynman diagrams and $S$-matrix amplitudes, to the concept of tunnelling through an external potential, demonstrating equivalence in neutral scalar systems. 

The choice of a delta function potential to investigate tunnelling is dually motivated by its computational ease, and by the fact a delta function is a useful approximation to a rectangular barrier. Indeed, it is possible to approximate a barrier with an arbitrary number of delta functions. Therefore, an important question to consider is when direct substitution of a Dirac delta potential can reproduce the correct limit of a rectangular barrier or other sharply-peaked functions. In non-relativistic quantum mechanics, directly solving the Schr\"odinger equation with a delta function reproduces the expected limit of a rectangular barrier. This too is the case for  transmission/reflection amplitudes obtained via the Klein-Gordon equation, solved with a Dirac delta potential. However, as highlighted by Subramanian and Bhagwat~\cite{subramanian_1971} in their relativistic treatment of the Saxon-Hutner theorem, solving the Dirac equation with a Dirac delta potential does not reproduce the equivalent limit from the solution of a rectangular barrier (the relativistic Kronig-Penney model). This apparent discrepancy is resolved by Calkin {\it et al}~\cite{calkinProperTreatmentDelta1987}, who argue that because the Dirac equation is a first-order PDE, a solution for a Dirac delta potential implies the spinor wavefunction is discontinuous at the boundary. Therefore, the canonical approach to obtain boundary conditions by integrating the Dirac equation over an infinitesimal region about the delta potential is ill-defined without further constraints on the delta potential. Through a more careful analysis, one finds the correct treatment is to consider a delta potential as a physical limit of a sharply-peaked potential, in which case there is no paradox. This subtlety warrants comment: performing the same Feynman diagram calculations with analytic continuation for the Dirac  Lagrangian coupled to an external delta potential, yields not the limit of a rectangular barrier, but rather the same result from a na\"ive integration about the delta function. This fact does not invalidate the entire summation process, but rather highlights the pathological problems with delta functions in fermion systems. 
\subsection*{Acknowledgements}
The authors would like to acknowledge  valuable discussions with A.G. Williams, and his close reading of the manuscript. This work has been supported by the Australian Research Council Discovery Project (Project No. DP190100256). R.Z. acknowledges the support of the Australian National University through the Dunbar Physics Honours Scholarship. 

\bibliography{QFT_tunnelling_references_master}

\section*{METHODS}
\subsection*{Derivation of tunnelling amplitude from Klein-Gordon equation and boundary conditions \label{app:KG_derivation}}
The quantum field theoretic approach to evaluate tunnelling amplitudes provides results that are identical to those computed from RQM for one and two Dirac delta potentials. 
The RQM derivations of these amplitudes are presented below.

\subsubsection{One-delta potential}
The transmission and reflection coefficients for a scalar field tunnelling through a Dirac delta barrier are derived directly using the Klein-Gordon equation:
\begin{equation}
(\partial^2+m^2)\phi(x) = -eaV_0\delta(x_3)\phi(x).
\end{equation}
The delta function divides the $x_3$ axis into two regions, $x_3<0$ and $x_3>0$. The solutions in each region are simply the free-particle solutions:
\begin{align}
\phi(x) =&\,\, e^{-i(p_0t-p_1x_1-p_2x_2)}\times \nonumber\\
&\begin{cases}
A_- e^{-ip_3x_3}+A_+ e^{ip_3x_3} &\qquad x_3 <0\\
B_+ e^{ip_3x_3} &\qquad x_3 >0,\\
\end{cases}
\end{align}
where $+$ ($-$) index indicates propagation towards increasing (decreasing) values of $x_3$.
The incoming wave is incident from $x_3\rightarrow-\infty$.

From continuity at $x_3 = 0$, 
\begin{align}
A_- +A_+ &= B_+. \label{eqn:boundary_1}
\end{align}
Due to the discontinuous nature of the delta function, the derivative of $\phi(x)$ will not be defined at the barrier. However, we can integrate the original Klein-Gordon equation over a small region of $x_3 \in [-\varepsilon, \varepsilon]$, for $\varepsilon \to 0$ 
\begin{align}
\int^{+\varepsilon}_{-\varepsilon} \mathrm{d}x_3  \,(\partial^2+m^2)\phi(x)&= -\int^{+\varepsilon}_{-\varepsilon}\mathrm{d}x_3\,eaV_0\delta(x_3)\phi(x) \label{eq:funnyintegral}\\
\left[\partial_{x_3}\phi(x)\right]_{x_3 = -\varepsilon}^{x_3 = \varepsilon}&=eaV_0 \phi(x)\big|_{x_3=0}\\
ip_3(B_+-A_++A_-) &= eaV_0 B_+ \label{eqn:boundary_2},
\end{align}
which gives another boundary condition. Note, only the $\partial_{x_3}^2\phi(x)$ term gives a non-zero integral as $\epsilon\to 0$ on the lefthand side.  
Solving Eqs.~(\ref{eqn:boundary_1}) and~(\ref{eqn:boundary_2}) for the transmission coefficient $T = \frac{B_+}{A_+}$ and reflection coefficient $R = \frac{A_-}{A_+}$ gives
\begin{align}
T &= \frac{1}{\frac{iaeV_0}{2p_3}+1}\\
R & = \frac{1}{\frac{2ip_3}{aeV_0}-1},
\end{align}
exactly congruent to the results in the main text.

\subsubsection{Double-delta potential} 
Let us now solve for the transmission and reflection amplitudes for the double-delta potential, with the associated Klein-Gordon equation:
\begin{align}
\left(\partial^2+m^2\right)\phi= -\frac{eaV_0}{2}\left[\delta\left(x_3+\frac{b}{2}\right)+\delta\left(x_3-\frac{b}{2}\right)\right]\phi.
\end{align}
This yields free-particle solutions in three regions of space:
\small
\begin{align}
\phi(x) = &\,\, e^{-i(p_0t-p_1x_1-p_2x_2)}\times \nonumber\\
&\begin{cases}
A_- e^{-ip_3x_3}+A_+ e^{ip_3x_3} \qquad &x_3 <-\frac{b}{2}\\
B_- e^{-ip_3x_3}+B_+ e^{ip_3x_3} \qquad -\frac{b}{2} < &x_3 <\frac{b}{2}\\
C_+ e^{ip_3x_3} \qquad &x_3 >\frac{b}{2}.\\
\end{cases}
\end{align}
\normalsize
There is only an outgoing wave in the region $x_3 >\frac{b}{2}$, corresponding to the transmitted wave. Continuity at each boundary gives: 
\begin{align}
A_-e^{ip_3b}+A_+ &= B_-e^{ip_3b}+B_+ \label{eqn:BCdouble_1}\\
B_-e^{-ip_3b}+B_+ &= C_+. \label{eqn:BCdouble_2}
\end{align}
Again, we cannot use continuity of the first derivatives to get another condition. We use the same integral as in Eq.~(\ref{eq:funnyintegral}) about a small region of each barrier to obtain:
\begin{align}
ip_3\left (A_-e^{ip_3b}-A_+  -B_-e^{ip_3b}+B_+\right )&= \nonumber\\
\frac{eaV_0}{2}\left (B_-\right.&\left.e^{ip_3b}+B_+\right ),\label{eqn:boundary_double_1}\\
ip_3(C_+-B_++B_-e^{-ip_3b})&= \frac{eaV_0}{2}C_+. \label{eqn:boundary_double_2}
\end{align}
Solving the system of equations given by Eqs.~(\ref{eqn:BCdouble_1}), (\ref{eqn:BCdouble_2}), (\ref{eqn:boundary_double_1}) and~(\ref{eqn:boundary_double_2}) gives
\begin{align}
    T_{2 \delta}  =&\frac{C_+}{A_+}= \frac{16 p_3^2 }{(a e V_0e^{i b p_3})^2+ (4 p_3+ ia e V_0)^2}\\
    R_{2 \delta} =&\frac{A_-}{A_+}= -\frac{2 i a e V_0 (a e V_0 \sin (b p_3)+4 p_3 \cos (b p_3))}{(a e V_0e^{i b p_3})^2+ (ia e V_0+4 p_3)^2}.
    \end{align}

\onecolumngrid
\subsection*{Derivation of tunnelling amplitude from $S$-matrix with a double Dirac delta potential}
The Fourier transform of the double Dirac delta potential defined in Eq.~(\ref{eq:pot2delta}) is
\begin{align}
\tilde{u}(q) &= (2\pi)^3 a V_0\cos\left (\frac{bq_3}{2}\right ) \prod_{\mu=0}^{2}\delta(p_\mu-k_\mu).
\end{align}
We take an initial state with momentum $p = (p_0, 0, 0, p_3)$  and consider the scattering probability to some state with momentum $k$. 
The zeroth and first-order $S$-matrix elements are given by Eqs.~(\ref{eqn:S_0}) and~(\ref{eqn:S_1}) respectively. To compute the second-order element, we use Eq.~(\ref{eqn:Sn_recursive}) and find
\begin{align}
\langle k | S | p \rangle ^{(2)} =& (-ie a V_0)^2(2\pi)^2\prod_{\mu=0}^{2}\delta(p_{\mu}-k_{\mu})\int \mathrm{d}q_3 \frac{i\cos\left (\frac{b}{2}(p_3-q_3)\right )\cos\left (\frac{b}{2}(q_3-k_3)\right ) }{p_3^2-q_3^2+i\varepsilon}\nonumber\\
=& \frac{ (-ia e V_0)^2}{4p_3} (2\pi)^3 \prod_{\mu=0}^{2}\delta(p_{\nu}-k_{\nu})\left[e^{i b p_3}\cos \left(\frac{b}{2} (k_3+p_3)\right)+ \cos
   \left(\frac{b}{2}  (p_3-k_3)\right)\right], \label{eqn:S_2_2_delta}
\end{align}
\normalsize
where we have used the on-shell condition $p_0^2-m^2 = p_3^2$ together with the property
  \begin{equation}
    \int \mathrm{d}q_3 \frac{\cos\left (\frac{b}{2}(p_3-q_3)\right )\cos\left (\frac{b}{2}(q_3-k_3)\right ) }{p_3^2-q_3^2+i\varepsilon} = -\frac{i \pi}{2p_3}  \left[\cos \left(\frac{b}{2} (k_3-p_3)\right)+e^{i b p_3} \cos \left(\frac{b}{2} (k_3+p_3)\right)\right]. \label{eqn:intneg}
    \end{equation}
The third-order element can be found analogously through a substitution of Eq.~(\ref{eqn:S_2_2_delta}) into Eq.~(\ref{eqn:Sn_recursive}), giving
  \begin{align}
    &\langle k | S | p \rangle ^{(3)} = \frac{(- iea V_0)^3}{8\pi p_3}(2\pi)^3 \prod_{\mu=0}^{2}\delta(p_{\mu}-k_{\mu})\int \mathrm{d}q_3\frac{ i\cos \left(\frac{b}{2} (k_3-q_3)\right) \left(\cos \left(\frac{b}{2}
       (p_3+q_3)\right)e^{ibp_3}+ \cos \left(\frac{b}{2} (q_3-p_3)\right)\right)}{p_3^2-q_3^2+i \epsilon}\nonumber\\
     &\quad\quad  =\frac{(-ia e V_0)^3}{16p_3^2} (2\pi)^3 \prod_{\mu=0}^{2}\delta(p_{\mu}-k_{\mu})\left[ \cos \left(\frac{b}{2}
       \left(k_3-p_3\right)\right)(1+e^{2ibp_3})+2 e^{i b p_3} \cos \left(\frac{b}{2}
       \left(k_3+p_3\right)\right)\right],
    \end{align}
where we have used Eq.~(\ref{eqn:intneg}).
Similarly, the expression for the $4^{th}$-order $S$-matrix element  reads
  \begin{align}
    &\langle k | S | p \rangle ^{(4)} = \frac{(-ia e V_0)^4}{64p_3^3} (2\pi)^3 \prod_{\mu=0}^{2}\delta(p_{\mu}-k_{\mu})\left[ \cos \left(\frac{b}{2}
       \left(k_3-p_3\right)\right)(1+3e^{2ibp_3})+\cos \left(\frac{b}{2}
       \left(k_3+p_3\right)\right)(3e^{ibp_3}+e^{3ibp_3})\right].
    \end{align}
These expressions quickly become unwieldy, though we can generalise the result to an arbitrary $n^{\text{th}}$-order term without knowledge of the previous term. Firstly, notice that the momentum difference term, $\cos \left(\frac{b}{2}
\left(k_3-p_3\right)\right)$ picks out the even terms in the binomial expansion $(1+e^{ibp_3})^n$, while the momentum sum term $\cos \left(\frac{b}{2}
\left(k_3+p_3\right)\right)$ will get the odd terms.  The $n^{th}$-order ($n\geq 1$) $S$-matrix element can then be written as
  \begin{align}
    \langle k | S | p \rangle ^{(n)} =&\frac{(-i ea V_0)^n}{2(4p_3)^{n-1}}(2\pi)^3 \prod_{\mu=0}^{2}\delta(p_{\mu}-k_{\mu})\left \{\cos \left(\frac{b}{2}
    \left(k_3-p_3\right)\right)\left [(1+e^{ibp_3})^{n-1}+(1-e^{ibp_3})^{n-1}\right ]\nonumber \right . \\
    & \left. + \cos \left(\frac{b}{2}
    \left(k_3+p_3\right)\right)\left [(1+e^{ibp_3})^{n-1}-(1-e^{ibp_3})^{n-1}\right ]\right \}.
    \end{align}
The sum over all orders can be found as a sum of four geometric series, yielding
  \begin{align}
\langle k | S | p \rangle  =& \langle k |S |p \rangle ^{(0)} + 4 eaV_0 p_3 (2\pi)^3 \prod_{\mu=0}^{2}\delta(p_{\nu}-k_{\nu})\frac{ (a e
       V_0-4 i p_3) \cos \left(\frac{b}{2} (k_3-p_3)\right)-e^{i b p_3}a e V_0 \cos \left(\frac{b}{2} (k_3+p_3)\right)}{(ia e V_0+4 p_3)^2+(a e V_0e^{ibp_3})^2}.
    \end{align}
Using Eqs.~(\ref{eqn:trans_defn}) and~(\ref{eqn:ref_defn}) gives the resulting transmission and reflection amplitudes in Eqs.~(\ref{eq:T2}) and~(\ref{eq:R2}).

\end{document}